\documentclass[aps,prl,twocolumn,groupedaddress,showpacs,amsmath,amssymb]{revtex4}

\usepackage{graphicx}

\begin{document}


\title{Phase Diagram and Spin Dynamics in Volborthite with a Distorted Kagom\'{e} Lattice}


\author{M. Yoshida}
\author{M. Takigawa}
\author{H. Yoshida}
\author{Y. Okamoto}
\author{Z. Hiroi}
\affiliation{Institute for Solid State Physics, University of Tokyo, Kashiwa, Chiba 277-8581, Japan}


\date{\today}

\begin{abstract}
We report $^{51}$V-NMR study on a high-quality powder sample of volborthite 
Cu$_3$V$_2$O$_7$(OH)$_2 \cdot  $2H$_2$O, a spin-1/2 Heisenberg antiferromagnet 
on a distorted kagom\'{e} lattice formed by isosceles triangles. In the magnetic fields 
below 4.5~T, a sharp peak in the nuclear spin-lattice relaxation rate 1/$T_1$ accompanied 
with line broadening revealed a magnetic transition near 1~K. The low temperature 
phase shows anomalies such as a Lorentzian line shape, a 1/$T_1 \propto T$ behavior 
indicating dense low energy excitations, and a large spin-echo decay rate 
1/$T_2$ pointing to unusually slow fluctuations. Another magnetic phase appears 
above 4.5~T with less anomalous spectral shape and dynamics. 
\end{abstract}


\maketitle



The search for exotic ground states in two-dimensional (2D) spin systems with frustrated 
interactions has been a challenge in condensed matter physics \cite{Misguich}. In particular, 
the ground state of the spin-1/2 Heisenberg model with a nearest neighbor interaction on a 
kagom\'{e} lattice, a 2D network of corner-sharing equilateral triangles, is believed to show 
no long-range magnetic order. Theories have proposed various ground states such as 
spin liquids with no broken symmetry with \cite{Waldtmann} or without \cite{Hermele} a spin-gap  
or symmetry breaking valence-bond-crystal states \cite{Singh}. 
Candidate materials known to date, however, 
depart from the ideal kagom\'{e} model in one way or another such as disorder, structural distortion, 
anisotropy, or longer range interactions. Volborthite Cu$_3$V$_2$O$_7$(OH)$_2 \cdot  $2H$_2$O 
is an example, which has distorted kagom\'{e} layers formed by isosceles 
triangles. Consequently, it has two Cu sites and two kinds of exchange interactions as shown 
in the inset of Fig.~1. The magnetic susceptibility $\chi$ obeys the Curie-Weiss law $\chi  = C/(T + \theta _W)$ above 
200~K with $\theta _W$ = 115~K, exhibits a broad maximum at 20~K, and 
approaches a finite value at the lowest temperatures, indicating absence of a spin gap. 
The specific heat and $\chi$ data revealed no magnetic order down to 2~K, 
much lower than $\theta _W$ \cite{Hiroi}. This indicates strong effects of frustration 
common to the geometry of the kagom\'{e} lattice, even though the 
difference between $J$ and $J^{\prime}$ should partially lift the massive 
degeneracy of low energy states of the ideal kagom\'{e} model. 

\begin{figure}[b]
\includegraphics[width=7.5cm]{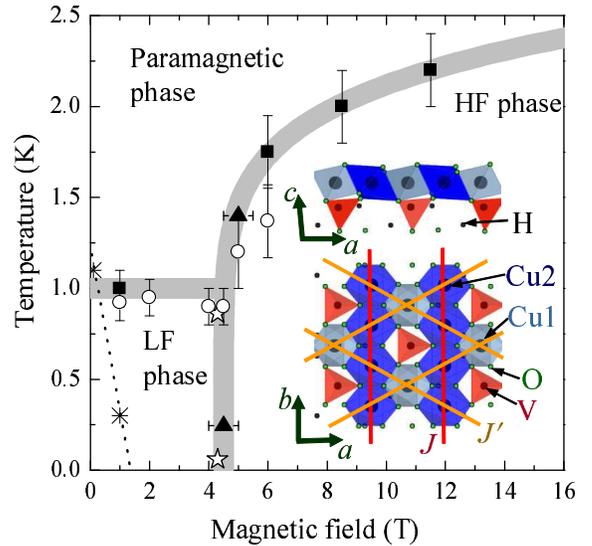}
\caption{\label{fig:fig1} (color online) Inset: the crystal structure of volborthite. 
Main panel: the proposed phase diagram. 
The squares (triangles) represent the phase boundaries determined 
from the broadening of NMR spectra as a function of temperature 
(magnetic field). The circles and asterisks indicate the 
peak of 1/$T_1$ and the onset of hysteresis in $\chi $, respectively. The stars 
indicate the first magnetization step \cite{Yoshida}. 
The lines are guide to the eyes.}
\end{figure}

Dynamic measurements, however, suggest a magnetic transition 
at a lower temperature. The nuclear spin-lattice relaxation rate 1/$T_1$ at the V sites 
shows a peak at 1.4~K, below which the nuclear magnetic resonance (NMR) spectrum 
begins to broaden \cite{Bert1}. The muon spin relaxation ($\mu$SR) experiments also 
detected slowing down of spin fluctuations with decreasing temperature towards 1~K \cite{Fukaya}. 
These dynamic anomalies coincide with the appearance of hysteresis in $\chi $, i.e. the difference 
between field-cooled and zero-field-cooled magnetization, suggesting a spin-glass like 
state \cite{Bert1}. However, impurity effects corresponding to the Curie term in 
$\chi$ of more than 0.5~\%/Cu of spin-1/2 \cite{Hiroi,Bert2} have been impeding 
proper understanding of intrinsic properties of volborthite. 

Recently, H. Yoshida \textit{et al.} have succeeded in reducing impurities 
down to 0.07 \% by hydrothermal annealing \cite{Yoshida}. 
Although a small hysteresis of $\chi$ is still observed below 1~K at 0.1~T, it is rapidly 
suppressed by magnetic fields of less than 2~T (the dotted line in Fig.~1). They 
also reported unusual sequential magnetization steps at 4.3, 25.5, and 46~T. 
In this Letter, we report on the $^{51}$V-NMR experiments using a high-quality powder 
sample prepared by the same method. We observed a sharp peak of 
1/$T_1$ at 0.9~K accompanied by a broadening of the NMR spectrum in 
the field range 1~$-$~4.5~T, indicating a magnetic transition unrelated to 
the hysteresis of $\chi $. The results of nuclear relaxation rates, however, 
revealed a large weight of low-energy excitations 
and persistence of anomalously slow spin fluctuations down to $T$ = 0. We also found 
another magnetic phase above 4.5~T with distinct spin structure and dynamics. 
The transition occurs at the same field as the first magnetization step \cite{Yoshida}. 
 
The NMR spectra were obtained by summing the Fourier transform of the spin-echo 
signal obtained by the pulse sequence $\pi/2$$-$$\tau$$-$$\pi/2$ at equally spaced magnetic 
fields $B$ with a fixed resonance frequency $\nu$. Typically $\tau$ = 12$-$20~$\mu$s 
and the pulse width was 1.0$-$1.8~$\mu$s. 
We determined 1/$T_1$ by fitting the spin-echo intensity $M(t)$ as a function of 
the time $t$ after several saturating comb pulses to the exponential recovery function 
$M(t)$ = $M_{eq}$$-$$M_0\mathrm{exp}(-t/T_1)$, 
where $M_{eq}$ is the intensity at thermal equilibrium. 
When this function did not fit 
the data due to inhomogeneous distribution of 1/$T_1$, we used the 
stretched exponential function $M(t)$ = $M_{eq}$$-$$M_0\mathrm{exp}\{-(t/T_1)^{\beta }\}$ 
to determine the representative value of 1/$T_1$. 

\begin{figure}
\includegraphics[width=8cm]{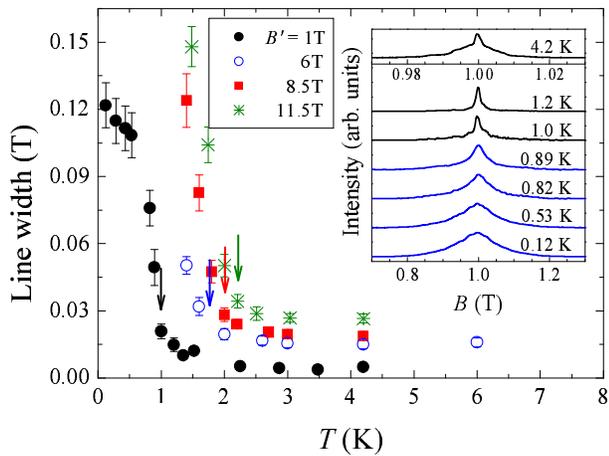}
\caption{\label{fig:fig2} (color online) $T$-dependence of the line width (FWHM) at various 
magnetic fields. ($B^{\prime}$ stands for the center of gravity of the spectrum.) 
The inset shows the NMR spectra at $\nu $ = 11.243~MHz ($B^{\prime} = 1$~T). The horizontal scale is expanded 
for the data at 4.2 K.}
\end{figure}

The inset of Fig.~2 shows the $T$-dependence of the $^{51}$V NMR spectrum 
below 4.2~K at $\nu$ = 11.243 MHz corresponding to $B$ $\sim$ 1~T. The spectrum 
at 4.2~K consists of a sharp central line with the full width at half maximum (FWHM) of 
$4 \times 10^{-3}$~T and quadrupole satellite lines (FWHM = $2 \times 10^{-2}$~T). 
At high temperatures above 140~K, the magnetic 
hyperfine shift $K$ and $\chi$ follow the same $T$-dependence, yielding the hyperfine 
coupling constant $A_{hf} = 0.77$~T/$\mu _B$ consistent with the previous 
reports \cite{Hiroi,Bert1}. The FWHM of the central line is about a half of the shift, indicating 
that the hyperfine coupling is dominantly isotropic, which should be attributed to the neighboring 
six Cu$^{2+}$ spins. Below 4.2~K, gradual broadening of the central line smears out the 
quadrupole structure. In the temperature range 2 $-$ 4~K, the width of the whole spectrum 
is about a factor four smaller than the previous results \cite{Hiroi,Bert1}, indicating much 
less disorder in our sample. 

The spectrum shows a sudden broadening below 1~K and the width tends to saturate 
below 0.5~K, where the line shape is approximately Lorentzian. 
The $T$-dependence of the line width (FWHM of the whole spectrum) 
is shown in Fig.~2. Since the width at low temperatures is independent 
of $B$ below 3.5~T as will be shown later (Fig.~4), the symmetric line broadening must be 
due to spontaneous staggered moments, which remain finite at zero field. However, the line 
shape does not look like that of an ordinary antiferromagnet. For a powder 
sample of an antiferromagnet with a homogeneous magnitude of moments, random orientation of the 
internal field with respect to the external field yields a rectangular spectral shape. The Lorentzian 
line shape in volborthite indicates large distribution of the magnitude of moments, suggesting 
a spin-density-wave like modulation or a lack of long range spatial order. Note 
that the transition near 1~K is \textit{not} a spin-glass transition reported in the earlier work \cite{Bert1}, 
since the hysteresis in $\chi$ in our sample is barely visible only below 0.3~K at $B = 1$~T (Fig.~1 \cite{Yoshida}). 

\begin{figure}
\includegraphics[width=8cm]{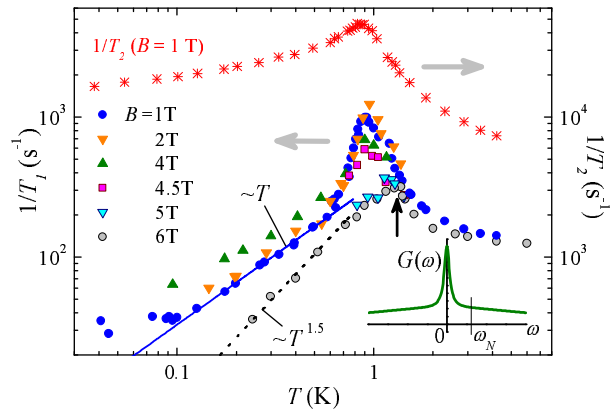}
\caption{\label{fig:fig3} (color online) $T$-dependence of 1/$T_1$ at various magnetic fields $B$. 
The asterisks represent the $T$-dependence of 1/$T_2$ at 1~T. The solid (dotted) line shows 
the power law dependence $T^\alpha$ with $\alpha $ = 1 (1.5). 
The inset shows a schematic illustration of $G(\omega )$ with two distinct frequency scales.}
\end{figure}

Figure~3 shows the $T$-dependence of 1/$T_1$ measured at the peak of the spectra 
for various values of $B$. 
One can recognize an apparent change of behavior near $B$ = 4.5~T. We first 
discuss the results below 4.5~T. We were able to fit $M(t)$ to the single exponential 
function above 1.2~K. Below 1.2~K, however, distribution in 1/$T_1$ let us 
use the stretched exponential function. The stretch exponent $\beta $ decreased monotonically
with decreasing $T$ and stayed constant at $\beta$ = 0.6 below 0.8~K down to 40~mK. 
A sharp peak of 1/$T_1$ is observed at 0.9~K for $B$ = 1~T, which 
agrees with the onset of line broadening, providing further evidence for 
a magnetic transition. The transition temperature of 1~K is not far from that 
reported by Bert \textit{et al.} (1.4~K) \cite{Bert1}. However, our data show a much 
sharper peak and the peak value of 1/$T_1$ is nearly an order of magnitude larger 
than their value. Thus the transition becomes more pronounced when disorder is reduced. 
The peak temperature of 1/$T_1$ is nearly independent of $B$ below 4.5~T. 
This behavior is again in contradiction to the hysteresis in $\chi$ or the spin-glass-like behavior. 

At low temperatures below 0.6~K, 1/$T_1$ varies proportionally to $T$. 
While the magnetic field enhances 1/$T_1$ slightly, the 
$T$-linear behavior is robust up to 4~T. Below 0.1~K, 1/$T_1$ 
becomes independent of $T$, which is most likely caused by impurities. 
The $T$-linear behavior is in strong contrast to what is observed in ordinary 
antiferromagnets, where nuclear relaxation is caused by the scattering of 
spin waves. Usually the two-magnon Raman process is dominant for 
anisotropic hyperfine coupling, while three-magnon process has to be 
invoked for isotropic coupling \cite{Moriya1, Moriya2, Beeman}. For $T \gg \Delta$, 
where $\Delta$ is the gap in the spin wave spectrum, 1/$T_1 \propto T^D$ \cite{Moriya1,Mila} 
for the two-magnon process, where $D$ (= 3 or 2) is the dimensionality of the spin-wave dispersion. 
The three magnon process also leads to power laws with even larger exponents. 
In both dimensions, 1/$T_1$ decreases exponentially for $T \ll \Delta$. 
The $T$-linear behavior without any sign of a spin gap in volborthite, therefore, indicates 
anomalously dense low energy excitations at low temperatures. 

A possible mechanism for the $T$-linear behavior of 1/$T_1$ is the particle-hole 
excitations in a fermionic system as in metals. Since fermionic 
elementary excitations (spinons) are proposed for the kagom\'{e} systems \cite{Hermele,Hao}, 
it is appropriate to compare our data with the value for free fermions given by the 
Korringa relation, $1/ \left( T_1TK^2 \right) = 4 \pi k_B \hbar (\gamma_N/g\mu_B)^2$. 
From the value of the shift at 4.2~K ($K$ = 0.40~\%), we found that the free fermion value 
$1/ \left( T_1 T \right)$ = 4~sec$^{-1}$K$^{-1}$ is two orders of magnitude smaller than the 
experimental value 330~sec$^{-1}$K$^{-1}$. (Note that $\chi$ at 1~T is nearly independent 
of $T$ below 4~K \cite{Yoshida}.) Thus the dynamics involves much slower fluctuations 
than static energy scale set by $\chi$ or the spinon band width. 
Another possible origin is the localized zero-energy mode with nearly flat dispersion, as was 
observed in classical Kagom\'{e} systems \cite{Matan}. 
  
Quite surprisingly, even slower fluctuations were revealed from the 
spin-echo decay rate 1/$T_2$. Figure~3 also shows the 
$T$-dependence of 1/$T_2$ measured at $B$ = 1~T. The spin-echo 
intensity exhibits exponential $\tau$ dependence, $M(2\tau) \propto \mathrm{exp}(-2\tau /T_2)$. 
This means that the fluctuations of the local-field $h(t)$ causing 
the spin-echo decay are in the narrowing limit, $\gamma \sqrt{\langle h^2(t) \rangle} \tau_c \ll  1$, 
where $\tau_c$ is the correlation time. 
Above 0.9~K, 1/$T_2$ and 1/$T_1$ show similar $T$-dependence but largely 
different values $r = (1/T_2)/(1/T_1) \sim 50$. Below 0.9 K, they show qualitatively 
different behavior. 1/$T_2$ decreases much more slowly than 1/$T_1$
and approaches a large value towards $T$ = 0 ($r \sim 500$). 

In general, 1/$T_2$ is the sum of electronic and nuclear contributions, $1/T_2 = (1/T_2)_e + (1/T_2)_n$. 
The first term further consists of two terms due to fluctuations of the hyperfine field perpendicular and 
parallel to the external field, $(1/T_2)_e = (1/T_2)_{e\perp} + (1/T_2)_{e\parallel}$, 
where $(1/T_2)_{e\perp} = 1/(2T_1) = (\gamma^2/2) \int_{-\infty}^{\infty} \langle h_{\perp}(0) h_{\perp}(t) \rangle \exp(2 \pi i \nu t) dt$ 
is given by the fluctuation amplitude of $h_{\perp}(t)$ at the NMR frequency $\nu$ \cite{note1}.  
On the other hand, $(1/T_2)_{e\parallel}$ is determined by the longitudinal fluctuations averaged over the time 
scale of $T_2$ itself \cite{note2, Recchia}, i.e. sensitive to much slower fluctuations in the range of 1$-$10~kHz. 

The large value of $r$ requires either $(1/T_2)_n \gg  (1/T_2)_e$ or 
$ (1/T_2)_{e\parallel} \gg  (1/T_2)_{e\perp}$. The former 
case is excluded for several reasons. First, 1/$T_2$ due to the nuclear dipolar coupling 
is orders of magnitude smaller than the observed values. Second, we found that 1/$T_2$ at 0.25~K 
did not depend on the strength of the rf-field in spite of the large line width, proving 
that like-spin ($^{51}$V nuclei) contribution is negligible. If unlike-spins, $^{65, 63}$Cu or $^{1}$H, have 
strong indirect coupling to $^{51}$V nuclei, $\tau_c$ is nothing but $T_1$ of these unlike spins. 
Since 1/$T_2$ is proportional to $\tau_c$ in the narrowing limit, similar $T$-dependence 
of 1/$T_2$ and 1/$T_1$ of $^{51}$V would imply that $^{63,65}$Cu (or $^1$H) and 
$^{51}$V have completely opposite $T$-dependence of 1/$T_1$. This sounds extremely unlikely. 

Therefore, $ (1/T_2)_{e\parallel}$ $\gg$  $(1/T_2)_{e\perp}$ = $1/(2T_1)$. Since only 1~T of 
field should not be sufficient to induce strong anisotropy in a powder sample, 
we conclude that $ (1/T_2)_{e\parallel}$ is dominated by spin fluctuations 
with an extremely small frequency scale, much smaller than the NMR frequency ($\sim$ 10~MHz) 
but still large enough ($\ge$ 10~kHz) to make 1/$T_2$ of the order 
of 10$^{-4}$~sec$^{-1}$ in the narrowing limit. This implies that the spectral density 
$G(\omega )= \int_{-\infty}^{\infty} \langle h(0) h(t) \rangle \exp(i \omega t) dt$ has at least two frequency scales 
as illustrated in the inset of Fig.~3. One gives the narrow zero-frequency peak that determines 1/$T_2$ 
and another one characterizes the broad spectrum that determines 1/$T_1$. 

Now we discuss the 1/$T_1$ data above $B$ = 4.5~T (Fig. 3). 
We fit the data of $M(t)$ by the single 
(stretched) exponential function above (below) 1.6~K. The value of $\beta $ 
decreased from 0.8 at 1.5~K to 0.25 at 0.14~K. While the sharp 
peak at 0.9~K observed below 4.5~T is absent, a broader peak appears 
near 1.3~K (see the arrow in Fig.~3). At low temperatures, 1/$T_1$ at 6~T 
decreases more rapidly with decreasing temperature, approximately as $T^{1.5}$, 
compared with the data below 4.5~T. 

A significant change in the line shape is also observed near 4.5~T. Figure 4(a) shows the 
$B$-dependence of the NMR spectrum at 0.28~K. Below 
3.5~T, the spectrum is approximately a Lorentzian with a $B$-independent width. 
At 4~T, two shoulders develop on both sides of the spectrum as indicated by the arrows, 
which eventually grow as the edges of a rectangular spectrum above 5~T. The results of 1/$T_1$ 
and the line shape combined together provide convincing evidence for a field-induced 
transition near 4.5~T. The rectangular spectral shape in the high field phase is compatible 
with homogeneous magnitude of moments. We emphasize that 4.5~T is close to the field 
at which the first magnetization step was observed \cite{Yoshida}. 

\begin{figure}
\includegraphics[width=7.5cm]{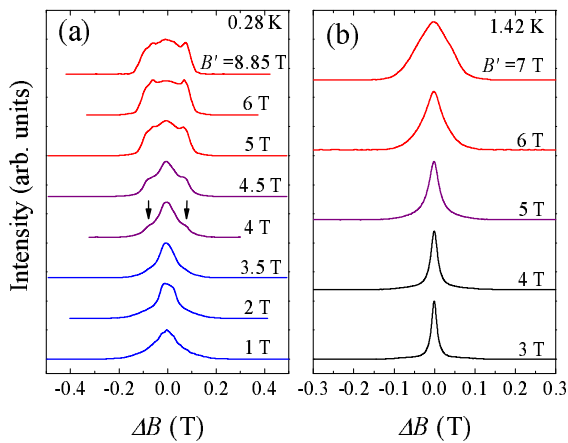}
\caption{\label{fig:fig4} (color online) Variation of the NMR spectrum with magnetic field at 
(a) $T$ = 0.28~K and (b) 1.42~K. The horizontal axis shows the deviation from the resonance field 
at 4.2 K, which represents the shift due to antiferromagnetic moments: $\Delta B = B-\nu/\gamma(1+K)$, 
where $\gamma$ and $K$ is the gyromagnetic ratio of $^{51}$V (11.1988~MHz/T) and 
the magnetic hyperfine shift at 4.2~K (0.4~\%), respectively.}
\end{figure}

In order to determine the phase diagram, we plot $T$-dependence of the line width 
at several fields in addition to the data at 1~T in Fig.~2. The onset of line 
broadening indicated by the arrows shifts to higher temperatures with increasing field. 
Figure 1 shows the $T$-$B$ phase diagram thus determined (solid squares). 
The peak temperatures of 1/$T_1$ are also displayed by the open circles. 
Three phases can be identified: the paramagnetic (P), low field (LF), and 
high field (HF) phases. It is remarkable that the transition temperature 
from P to LF phases is independent of $B$ up to 4~T. Note that there is no anomaly in 
the $T$-dependence of $\chi$ at 1~T \cite{Yoshida}. However, recent precise 
measurements of heat capacity revealed a weak anomaly near 1~K \cite{Yamashita} 
not detected before. In contrast, the transition between the P and HF 
phases moves to higher temperatures with increasing field. To further confirm the boundary 
between the P and HF phases, we examined the 
$B$-dependence of the spectrum at 1.4 K (Fig.~4b). The line width 
varies linearly with $B$ below 4~T, consistent with the paramagnetic 
behavior. However, it shows a sudden large increase above 5 T, supporting 
the transition directly from the P to HF phases. The transitions determined 
from the $B$-dependence of the spectra are shown by the triangles in Fig.~1. 
The HF phase appears less anomalous than the LF phase since the line shape 
is compatible with the ordinary antiferromagnets and the power law behavior of 1/$T_1$ 
is closer to the prediction of the spin wave theory in 2D. 

In conclusion, we found two magnetic phases in volborthite. The low field phase 
is anomalous in various aspects: a large variation in the magnitude of moments, 
dense low frequency excitations characterized by two distinct frequency scales, 
one of which is extremely small, and the transition insensitive to magnetic field. 
The high field phase appears less anomalous. Possible magnetic order has 
been discussed for distorted kagom\'{e} systems for the case of 
either $J/J^{\prime}>1$ \cite{Schnyder} or $J/J^{\prime}<1$ \cite{Wang}. 
Magnetic order in undistorted kagom\'{e} systems can be induced by 
the Dzyaloshinskii-Moriya interaction \cite{Cepas} or longer range 
interactions \cite{Domenge}. What causes the magnetic transition in volborthite 
remains an open issue. 

We thank H. Tsunetsugu and F. Mila for stimulating discussions and Y. Nakazawa for disclosing 
unpublished data. The work was supported by the G-COE program and by Grant-in-Aids for 
Scientific Research on Priority Areas "Novel States of Matter Induced by Frustration" (19052003) 
from MEXT Japan.


\begin{thebibliography}{99}
\bibitem{note2}The spin-echo decay due to fluctuations of $h_{\parallel}(t)$ is expressed as 
${M (2\tau)}/ {M(0)} = \langle \cos \left[ \gamma \left( \int_0^{\tau} h_{\parallel} (t) dt - \int_{\tau}^{2\tau} h_{\parallel} (t) dt \right) \right] \rangle$.
By assuming a Gaussian distribution for the accumulated phase, the argument of the cosine function, one obtains 
an exponential spin-echo decay in the narrowing limit \cite{Recchia}.
\end{thebibliography}
\end{document}